\documentclass[conference]{IEEEtran}

\AtBeginDocument{%
  }

\usepackage{makecell}
\usepackage{multirow}
\usepackage[utf8]{inputenc} % allow utf-8 input
\usepackage[T1]{fontenc}    % use 8-bit T1 fonts
\usepackage{threeparttable}
\usepackage{hyperref}       % hyperlinks
\usepackage{url}            % simple URL typesetting
\usepackage{booktabs}       % professional-quality tables
\usepackage{amsfonts}       % blackboard math symbols
\usepackage{nicefrac}       % compact symbols for 1/2, etc.
\usepackage{microtype}      % microtypography
\usepackage{subfigure}
\usepackage{soul}
\usepackage[table,xcdraw]{xcolor}
\usepackage{colortbl}

\usepackage{graphicx}
\usepackage{bbm}
\usepackage[linesnumbered,ruled,vlined]{algorithm2e}

\usepackage{caption}
\usepackage{subcaption}
\usepackage{textcomp}
\usepackage{amsmath}
\usepackage{balance}
\usepackage{wrapfig}

\newcommand{\model}{\textsc{scReader}}
\newcommand{\chillmodel}{\textsc{\textbf{LLM as \ul{S}ingle-\ul{C}ell RNA Data \ul{Reader}}}}
\newtheorem{definition}{Definition}
\renewcommand{\thefootnote}{}
\hypersetup{
colorlinks=true,
linkcolor=black
}
\begin{document}

\title{\model{}: Prompting Large Language Models to Interpret scRNA-seq Data}

\author{\IEEEauthorblockN{Cong Li$^{2,3,\dagger}$, Qingqing Long$^{1,\dagger}$, Yuanchun Zhou$^{1,3}$, Meng Xiao$^{1,*}$}
\IEEEauthorblockA{$^1$ Computer Network Information Center, Chinese Academy of Sciences\\
$^2$ State Key Laboratory of Stem Cell and Reproductive Biology, Institute of Zoology, Chinese Academy of Sciences\\
$^3$ University of Chinese Academy of Sciences\\
licong22@ioz.ac.cn, \{qqlong, zyc, shaow\}@cnic.cn
}}
\maketitle

\begin{abstract}
Large language models (LLMs) have demonstrated remarkable advancements, primarily due to their capabilities in modeling the hidden relationships within text sequences. 
This innovation presents a unique opportunity in the field of life sciences, where vast collections of single-cell omics data from multiple species provide a foundation for training foundational models. 
However, the challenge lies in the disparity of data scales across different species, hindering the development of a comprehensive model for interpreting genetic data across diverse organisms.
In this study, we propose an innovative hybrid approach that integrates the general knowledge capabilities of LLMs with domain-specific representation models for single-cell omics data interpretation. 
We begin by focusing on genes as the fundamental unit of representation. Gene representations are initialized using functional descriptions, leveraging the strengths of mature language models such as LLaMA-2. 
By inputting single-cell gene-level expression data with prompts, we effectively model cellular representations based on the differential expression levels of genes across various species and cell types.
In the experiments, we constructed developmental cells from humans and mice, specifically targeting cells that are challenging to annotate. We evaluated our methodology through basic tasks such as cell annotation and visualization analysis. The results demonstrate the efficacy of our approach compared to other methods using LLMs, highlighting significant improvements in accuracy and interoperability.
% This study showcases the potential of combining LLMs with biological datasets, paving the way for more nuanced and comprehensive models that can bridge the gap between general language understanding and specific biological interpretations. 
Our hybrid approach enhances the representation of single-cell data and offers a robust framework for future research in cross-species genetic analysis.\renewcommand{\thefootnote}{\arabic{footnote}}\footnote{Our code is publicly available via Github: \url{https://shorturl.at/ZIw02}.}
\end{abstract}
\renewcommand{\thefootnote}{}
\footnote{$^*$Meng Xiao (shaow@cnic.cn) is the Corresponding Author.}
\footnote{$^\dagger$These authors contributed equally to this study.}
\footnote{This work is partially supported by the Postdoctoral Fellowship Program of CPSF (No.GZC20232736) and the China Postdoctoral Science Foundation Funded Project (No.2023M743565), the Strategic Priority Research Program of the Chinese Academy of Sciences XDB38030300.}
\setcounter{footnote}{1}
\renewcommand{\thefootnote}{\arabic{footnote}}

\section{Introduction}
% Recent study, such as scGPT~\cite{cui2023scgpt}, scFoundation~\cite{hao2023scfoundation}, Geneformer~\cite{theodoris2023geneformer}, and GeneCompass~\cite{yang2023genecompass} 

% ~\cite{touvron2023llama}~\cite{ouyang2022training}
% Recent studies~\cite{cui2023scgpt,hao2023scfoundation,theodoris2023geneformer,yang2023genecompass} make significant progress in learning gene-level and cell-level representation by leveraging large-scale gene expression data. 

% However, there are few studies~\cite{chen2023genept} that focus on adopting the external description and common knowledge from Large Language Models 
%  (LLMs) to enhance the representation. 
% Despite the inherent limitations of existing LLMs~\cite{ye2023character}  in directly reading and interpreting gene expression data, they demonstrate significant potential and flexibility as the Foundation Model. 
% This research proposes \textbf{\model}, which focuses on utilizing the LLMs with the capability to directly interpret and distinguish cell types in gene expression data. 
% Our preliminary experiments indicate that \model excels in accurately categorizing known cell types, demonstrating the potential of the common knowledge in LLMs as an effective bridge for uncovering biological insights.

Large Language Models (LLMs)~\cite{achiam2023gpt, dubey2024llama} have revolutionized various fields by demonstrating exceptional capabilities in understanding and generating human language~\cite{zheng2024llamafactory}. 
Their ability to capture intricate patterns and relationships within sequential-like data makes them powerful tools for knowledge representation and natural language processing~\cite{lee2023benefits,ye2023needed,xiao2023hierarchical}. 
In recent years, LLMs have also shown potential in broader applications, such as interpreting structured data and providing insights across diverse domains~\cite{lecler2023revolutionizing}, such as bioinformatics~\cite{theodoris2023geneformer}, chemistry~\cite{castro2023large}, and scientometrics~\cite{cui2024automated,cai2023resolving}. 
This versatility stems from their extensive training on vast corpora, enabling them to encapsulate a wide range of general knowledge~\cite{zhou2020survey,dong2024temporal}. 
As Foundation Models, LLMs offer flexibility and scalability, making them well-suited for complex datasets, including those encountered in the life sciences~\cite{labrak2024biomistral,wang2024biorag}.
Despite these advancements, existing research in gene representation~\cite{chen2023genept} and cell-level analysis~\cite{theodoris2023geneformer} often fail to leverage the full potential of LLM. Studies \cite{cui2023scgpt,hao2023scfoundation,yang2023genecompass} have focused mainly on large-scale gene expression data but have not fully incorporated external descriptions and common knowledge available through LLM. 
This oversight limits the depth and richness of gene representations, as current models do not adequately utilize the comprehensive knowledge embedded in LLMs. 
Furthermore, the application of LLMs in life sciences remains underexplored, lacking opportunities to enhance the interpretation of biological data with linguistic insights.

In summary, the integration of large language models (LLMs) into genomic data interpretation faces several challenges. 
\textbf{(C1) Ignore Existing Knowledge:} There is insufficient utilization of detailed gene knowledge, such as the well-constructed description of gene functions~\cite{federhen2012ncbi,schoch2020ncbi}, which limits the enrichment of gene representations. 
\textbf{(C2) Limited Domain Data:} 
Although sequencing techniques generate numerous high-throughout data~\cite{schwartzman2015single,gawad2016single,woodworth2017building,zhang2024enhanced}, the disparity in data volume across species poses a challenge, as many species lack the extensive data required for training large models, complicating the development of universally applicable representations.
\textbf{(C3) Biomodel Semantic Gap:} Lastly, existing biological large language models are mostly trained solely on omics data, which is a type of sequence data, neglecting the vast amount of high-quality textual semantic information that has made large language models successful~\cite{dale2021gpt}. This approach causes existing models to overlook the common sense understanding of the world that humans possess~\cite{floridi2020gpt}.

To address these challenges, we propose a novel framework named \chillmodel{} (\textbf{\model}), which integrates LLMs with gene expression data interpretation.
Our strategy involves using functional gene descriptions to initialize gene representations, enhancing the model with detailed biological insights. 
By treating genes as the fundamental unit of analysis, we connect different species through common genetic knowledge, overcoming data limitations. 
In addition, we employ prompt learning techniques to leverage mature LLMs, harnessing their capabilities to interpret and distinguish cell types in gene expression data effectively.
Our contributions can be summarized as follows:
\begin{itemize}
    \item We introduce a method for initializing gene representations using functional descriptions, enriching the interpretative depth of genomic data.
    \item We propose a cross-species approach using genes as the basic unit, facilitating insights even with limited data availability.
    \item We demonstrate the application of prompt learning to adapt LLMs for life sciences, enhancing the analysis of gene expression data.
    \item Our preliminary experiments show that \textbf{\model} excels in accurately categorizing cell types, highlighting the utility of LLMs in bridging linguistic and biological knowledge to uncover new biological insights.
\end{itemize}

\section{Preliminary}
\subsection{Important Definitions}
\begin{definition}[\textbf{Gene Description}]
A gene description is a textual sequence that provides a concise summary of a gene's function, structure, and other relevant biological information. 
It typically includes details such as the gene's name, associated protein products, cellular localization, and its role in biological processes or pathways. 
Formally, a gene description can be denoted as a word sequence: $T = \{t_0, t_1, \cdots\}$.
\end{definition}

\begin{definition}[\textbf{Large Language Model}] 
A Large Language Model (LLM) can be conceptualized as a function that takes a word sequence as input and output: $f: \mathbf{R}^{n\times1} \rightarrow \mathbf{R}^{m\times1}$, where $n$ and $m$ are the length of the input and output sequence. 
\end{definition}

\begin{definition}[\textbf{Single-cell RNA sequencing Data}] scRNA-seq data can be represented as a sequence $C\in \mathbf{R}^{n_g\times 1}$, where $n_g$ is the total gene number. Each column of $C$ corresponds to a gene. 
The values in the matrix represent gene expression levels, typically measured in counts or normalized units.
\end{definition}

\subsection{Problem Statement} 
This study aims to evaluate the ability of LLM to understand and analyze single-cell RNA sequencing (scRNA-seq) data by focusing on the cell-type annotation task.
Cell type annotation is a crucial step in single-cell genomic analysis, involving the assignment of cell type labels to individual cells based on their gene expression profiles. 
Formally, the cell type annotation task can be defined as follows:
\begin{definition}[\textbf{Cell Type Annotation Task}]
Given A set of cells $C = \{c_1, c_2, \ldots, c_n\}$ and a set of predefined cell type labels $L = \{l_1, l_2, \ldots, l_k\}$
The task is to find a function $f: C \rightarrow L$ that assigns a cell type label to each cell, such that $f(c_i) = l_j$, where $l_j \in L$ is the most appropriate cell type label for cell $c_i$ based on its gene expression profile.
\end{definition}

\section{Related Work}
\noindent The transcriptome~\cite{martin2011next}, which refers to the collective expression states of thousands to tens of thousands of genes in each cell in biology~\cite{mutz2013transcriptome}, is a crucial determinant of the cell state. 
With the continuous advancement of modern sequencing technologies~\cite{scRNAseq_overview}, the number of transcriptomes~\cite{international2001initial,gao2018tracing} worldwide is accumulating astonishingly. 
Effectively interpreting this information and further analyzing life processes has become a significant challenge. 
Over the centuries of life science development, humans have gained a substantial understanding of life systems, including elucidating the functions of many genes~\cite{federhen2012ncbi, schoch2020ncbi}. 
This knowledge is stored in the medium of natural language. 
Although directly using LLMs to understand omics data might be quite challenging due to the significant differences between characters and gene meanings or the dimension curse~\cite{kiselev2019challenges(dimension_curse)}, initializing each gene based on the semantics of its function. 
At the same time, LLM offers a promising and feasible approach~\cite{chen2023genept}.
Previous works have established foundational models~\cite{szalata2024transformers} for directly interpreting transcriptome information. By adopting an approach similar to language models~\cite{kenton2019bert,kalyan2023survey}, these methods~\cite{cui2023scgpt, hao2023scfoundation} encode the categories and expression values of active genes into token embeddings, which are further applied with attention mechanisms and can capture the interactions between genes and their determinative roles in the overall state of the cell. 
The success of such foundation models also prompts new considerations: \textbf{could larger parameters and more knowledge-rich LLM be adapted to understand omics data as well?}
In this study, we introduce a novel approach to answer this question. 

% \section{Feature Transformation Framework}
\begin{figure*}
\centering
\includegraphics[width=\textwidth]{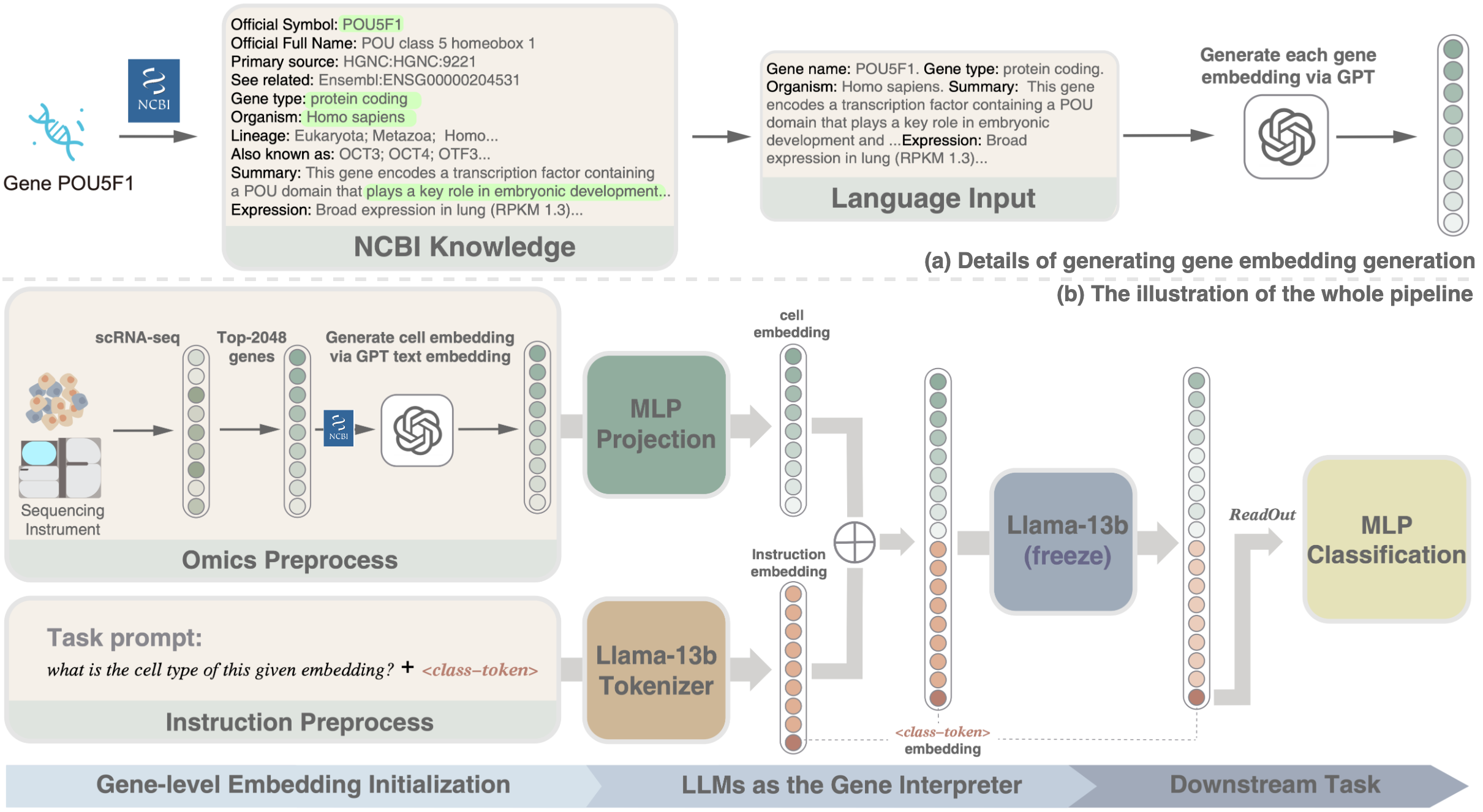}
\caption{The illustration of \model{}. (a) Details of generating gene embedding via NCBI gene description. (b) The pipeline of scInterpreter. The model will first embed each input from the cell and downstream task-specific instruction. Then, the cell embedding and instruction embedding will pass through the LLMs. After aggregating the knowledge and structural information of the given cell by LLMs, the model \textit{ReadOut} the representation and then conducts the downstream task.}
% \vspace{-0.3cm}
\label{main}
\end{figure*}

\section{Methodology}
As depicted in Figure~\ref{main}, \model{} consists of two main parts. 
The first part aims to generate the gene-level representation via LLMs based on the descriptive text of each gene.
The second part aims to generate the cell embedding via the specific gene expression and LLM.
After these stages, we could obtain the cell embedding and then feed it into the downstream task, such as cell-type annotation. 

\smallskip
\noindent\textbf{\ul{Gene-level Embedding Initialization.}}
The representation of each gene is initialized with its descriptive texts, which is extracted from the NCBI dataset\footnote{\url{https://www.ncbi.nlm.nih.gov/gene}}.
Those texts will then be fed into the GPT-3.5\footnote{\url{https://api.openai.com/v1/embeddings}} to generate the embeddings for each gene.

% \noindent\textit{\ul{Gene Description as Embedding.}} 

For each gene in our dataset, we extract its descriptive text from the NCBI Gene database. This database integrates information from a wide range of species and includes crucial details such as nomenclature, Reference Sequences (RefSeqs), maps, pathways, variations, and phenotypes. The rich textual information provided by NCBI serves as a comprehensive representation of each gene's biological context and function.
As illustrated in Figure~\ref{main}(a), we employ the GPT-3.5 model's embedding functionality to convert these textual descriptions into numerical representations suitable for computational analysis, defined as:
\begin{equation}
    e_i = f_{gpt}(T_{i}),
\end{equation}
where $T_{i}$ is the description of a given gene $g_i$, $f_{gpt}(\cdot)$ represents the \textit{text-embedding-ada-002} model in GPT-3.5, and $e$ is the representation of this gene. 

The process is as follows.
\begin{enumerate}
    \item \textbf{Data Extraction:} We query the NCBI Gene database using each gene's identifier or symbol to retrieve its associated descriptive text.
    
    \item \textbf{Text Preprocessing:} The extracted text is cleaned and standardized to ensure consistency across all gene descriptions. This may involve removing special characters, standardizing formatting, and truncating to a uniform length if necessary.
    
    \item \textbf{Embedding Generation:} Each preprocessed gene description is then passed through the GPT-3.5 embedding API. This API transforms the textual input into a high-dimensional vector (typically 1536 dimensions for GPT-3.5) that captures the semantic content of the description.
    
    \item \textbf{Embedding Storage:} The resulting embeddings are stored in a format that allows efficient retrieval and manipulation during subsequent analysis steps.
\end{enumerate}

This approach leverages the power of large language models to capture complex biological information in a dense, numerical format. 
By using GPT-3.5 for embedding generation, we benefit from its deep understanding of language and context, which has been trained on a vast corpus of text, including scientific literature. 
This allows our gene-level representations to potentially capture nuanced relationships and functional similarities that might not be immediately apparent in the raw text descriptions.

The resulting gene embeddings serve as the foundation for our subsequent analysis, providing a rich, contextual representation of each gene that goes beyond simple sequence-based or keyword-based approaches.

\smallskip
\noindent\textbf{\ul{Cell-level Representation.}}
Gene expression levels are crucial for distinguishing cell types in single-cell RNA sequencing data. 
We leverage this information by ranking genes based on their expression levels within each cell, combining this ordinal information with gene embeddings to construct a comprehensive cell representation. The process is as follows:

\begin{enumerate}
    \item \textbf{Expression Ranking:} For each cell, we rank all genes based on their expression levels in descending order. Let $[g_1, g_2, ..., g_n]$ represent the ranked sequence of genes for cell $c$, where $g_i$ is the $i$-th highest expressed gene in the cell. 
    We then take the top-2048 genes to form the ranked gene sequence. 
    
    \item \textbf{Embedding Lookup:} For each gene $g_i$ in the ranked sequence, we retrieve its corresponding embedding vector $e_i$ generated in the previous step (Gene-level Embedding Initialization).
    
    \item \textbf{Position-aware Representation:} To incorporate both the gene's identity and its relative expression level, we combine the gene embedding with its rank information. For the $i$-th ranked gene, we compute a position-aware representation $p_i$ as follows:
    
    \[p_i = e_i \oplus \text{PE}(i)\]
    
    where $\oplus$ denotes vector concatenation and $\text{PE}(i)$ is a positional encoding vector that captures the gene's rank.
    
    \item \textbf{Cell Representation:} The final representation for cell $c$ is then constructed as the sequence of these position-aware gene representations:
    
    \[E_c = (p_1, p_2, ..., p_n)\]
\end{enumerate}

This Ranked Gene Sequence approach offers several advantages: (1) It preserves the relative expression levels of genes within each cell, which is often more informative and robust than absolute expression values. (2) It combines the rich semantic information captured in gene embeddings with the cell-specific expression patterns. (3) The resulting representation is invariant to technical factors that might affect absolute expression levels across cells or experiments. (4) It provides a fixed-length representation for each cell, regardless of the number of non-zero expressed genes, facilitating downstream analysis. (5) Finally and most importantly, our approach only utilizes the knowledge in gene summary and LLM, so training is unnecessary. 

By constructing cell representations in this manner, we create a rich, informative input for subsequent cell-type annotation tasks that captures both the cell-specific expression patterns and the broader biological context of each gene.
% The total number of unique gene types observed in the human dataset is 23,111, among which 593 genes (approximately 2.5\%) lack corresponding entries in NCBI. For the mouse dataset, a total of 27,443 unique gene types were identified, with 4,503 genes (approximately 16.5\%) not found in NCBI. 
% Regarding these genes without entries in NCBI, we initiated direct question-and-answer requests. These requests solicited descriptions for the genes in the form of: "\textit{Please provide a description for gene x}" to be utilized as the descriptive text for the subsequent generation of embeddings for these genes.

\smallskip
\noindent\textbf{\ul{LLMs as the Gene Interpreter.}} 
We select Llama-13b~\cite{touvron2023llama} as the based LLM. 
The cell representation then projects to 5120 dimensions through a multi-layer perceptron (MLP) containing to conform to the Llama-13b's input dimensions $h$:
\begin{equation}
    E_{cell} = MLP_p(E_c),
\end{equation}
where $MLP_p(\cdot)$ is the projection layer with the learnable parameter. 
$C \in \mathbf{R}^{n\times h}$ is the embedding of the given cell. 

We then pass the cell embedding matrix into the Llama-13b along with the downstream task instruction, such as \textit{`what is the cell type of this given embedding?'}. After that, we take the \textit{class-token} ($<cls>$) from the output and feed it into a trainable classification head:
\begin{align}
% T_{ins} \xrightarrow[]{LLM} 
    E_{ins} \oplus E_{cell} \oplus e_{cls} \xrightarrow[ReadOut]{LLM} \hat{e}_{cls}, \\
    \hat{y} = \text{Softmax}(MLP_{c}(\hat{e}_{cls})),
\end{align}
where $E_{ins}$ is the text embedding of the given instruction, $e_{cls}$ is the embedding of the class-token. After fed into LLM, scInterpreter will $ReadOut$ the output. 
For the cell-type annotation task, we set the $ReadOut$ operation as directly taking the class-token embedding from the output. 
$\hat{e}_{cls}$ is the class-token embedding after the aggregation within the LLM. $\hat{y}$ is the model prediction.
During the training process, the Llama model will be frozen. The cross-entropy loss will optimize the projection layer, the classification head, and the \textit{class-token} token's embedding layer.

\smallskip
\noindent\textbf{\ul{Optimization Objective.}}
The model is trained to minimize the cross-entropy loss between the predicted label distribution and the true label distribution. For a single cell, the loss is defined as:
\[L = -\sum_{i=1}^{K} y_i \log(\hat{y}_i)\]
where $K$ is the number of cell type classes, $y_i$ is the true probability of the cell belonging to class $i$ (usually a one-hot encoded vector), and $\hat{y}_i$ is the predicted probability for class $i$.

\section{Data Preparation and Experimental Settings}
To validate \model{}, we construct two scRNA-seq datasets. 
HUMAN-10k comprises 10,000 single-cell sequencing records with 61 different cell types, each having 23,111 genes records. 
MOUSE-13k comprises 13,000  records with 37 different cell types, each having 27,443 gene records.

\smallskip

\noindent\textbf{\ul{Datasets Preparation.}}
The preprocessing and the golden label generation of the above single-cell RNA sequencing dataset are conducted via Seurat~\cite{Seuratv3} in R~\cite{team2000r}. 
First, we performed quality control to filter out low-quality cells and genes based on gene expression counts and mitochondrial gene percentage.
We normalized the data using log-normalization and identified highly variable genes with the FindVariableFeatures function. 
The data was scaled with ScaleData and subjected to Principal Component Analysis (PCA) using the top 10 principal components. 
We projected the data into two dimensions with UMAP and conducted clustering with the FindNeighbors and FindClusters functions (resolution = 0.5).
Marker genes were identified through differential expression analysis (FindAllMarkers) for each cluster. 
The golden label of the Cell type annotation was then identified by comparing the expression of marker genes with known cell type-specific genes from databases such as PanglaoDB~\cite{franzen2019panglaodb} and CellMarker~\cite{zhang2019cellmarker}. 
% The final annotation was validated by visualizing the clusters with UMAP and confirming the biological relevance of the results.
After cell clustering, manual annotation of cell types is performed by specifying marker genes for different cell clusters. 
Only genes with a log fold change greater than 0.3, expressed in at least 30\% of cells, and positively regulated are considered.

\smallskip
\noindent\textbf{\ul{Cell Embedding Initialization.}}
For the initialization of gene knowledge, we first selected 23,111 human genes and 27,443 mouse genes based on the expression median of each gene in the above dataset. Subsequently, we retrieved the description information for each gene from NCBI. The information entries included: Official Symbol, Official Full Name, Gene Type, Organism (organs and tissues where the gene is mainly distributed), Lineage, Expression (expression values in major organs obtained from experimental sequencing, represented in RPKM), and Summary (description of the gene's functional distribution based on past knowledge and experience). The retrieved gene knowledge was embedded using the GPT-3.5 API's `text-embedding-ada-002`, with the resulting embeddings having a dimension of 1536.
Among the 23,111 human genes, 593 did not have corresponding NCBI entries (approximately 2.5\%). 
For these genes, we manually queried the GPT-3.5 conversational model: "\textit{Please provide a detailed description of the functions, distribution, and expression of human gene $g_x$ as much as possible}" using the GPT-generated responses as the knowledge for these genes, where $g_x$ is the gene without description. 
Similarly, we employed the same method for embedding generation for the 4,503 mouse genes out of the 27,443 (approximately 16.4\%) that lacked NCBI entries. 
After obtaining the initialized embeddings of individual human and mouse genes, we screened the expression values of genes in each cell within the dataset. By normalizing using the median expression value of each gene, we only selected the top 2048 highly variable genes. The embedding of each gene was multiplied by its expression value in that cell, and then the 2048 new embeddings were stacked. This resulted in the initial cell embedding for each cell.

\smallskip
\noindent\textbf{\ul{Training Workflow.}}
Our methods were implemented according to the following workflow: The initialized embeddings were fed into a Multilayer Perceptron (consisting of 5 layers, where the 1536-dimensional input is projected to 4096 dimensions to fit the input requirements of Llama-13b). 
Then these embeddings were passed through a frozen Llama-13b model, followed by the embedded text instruction \textit{'What is the cell type of this given embedding?'}. A \textit{<cls>} token was attached to the end of the sFinally, \textit{<cls>} states were read out of hidden output states and applied with another \textit{MLP} classifier head to do cell-type classification training.
For comparison, we removed the frozen Llama model as the GenePT group, the two \textit{MLP}s remained and were trained normally. 
The initialized cell embeddings are first passed through an up-projection \textit{MLP}, increasing the dimensionality to 4096. Subsequently, the embeddings of each gene within the cell are averaged and then directly fed into a classification \textit{MLP} for cell-type classification training.
In both sets of experiments, the human and mouse datasets were randomly split into a 10-fold validation split ratio to serve as training and testing sets, respectively. Both sets of experiments were trained for 10 epochs, using the same learning rate of 5e-5, batch size of 64, and 1000 warm-up steps.

\smallskip
\noindent\textbf{\ul{Baseline Methods.}} 
We select \textbf{GenePT}~\cite{chen2023genept} as the compared method. GenePT is a novel approach that leverages pre-trained language models for interpreting single-cell RNA sequencing data. The selection of GenePT as our baseline is motivated by several factors:

\begin{itemize}
    \item \textbf{Similar Conceptual Framework:} Like our approach, GenePT utilizes the power of large language models to process and understand gene expression data, making it a suitable candidate for comparison.
    
    \item \textbf{Proven Effectiveness:} GenePT has demonstrated strong performance in various single-cell analysis tasks, including cell type annotation, which aligns with our primary objective.
    
    \item \textbf{Adaptability:} The method is adaptable to different types of single-cell data and various downstream tasks, allowing a fair comparison across different experimental settings.
\end{itemize}
In our comparative analysis, we evaluated both GenePT and our proposed method on the same datasets, using identical train-test splits and evaluation metrics. This ensures a fair and rigorous comparison of the two approaches.
\section{Experiments}
In this section, we conducted the cell-type annotation task to illustrate the advantage of introducing the large language model to facilitate understanding of the single-cell omic data. 

\begin{figure}
\centering
\subfigure[HUMAN-10k]{
    \includegraphics[width=0.45\linewidth]{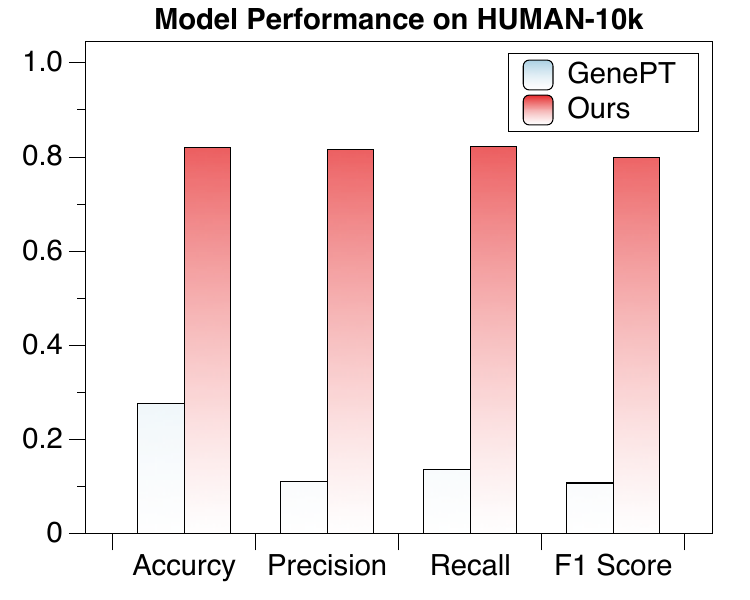}}
\subfigure[MOUSE-13k]{
    \includegraphics[width=0.45\linewidth]{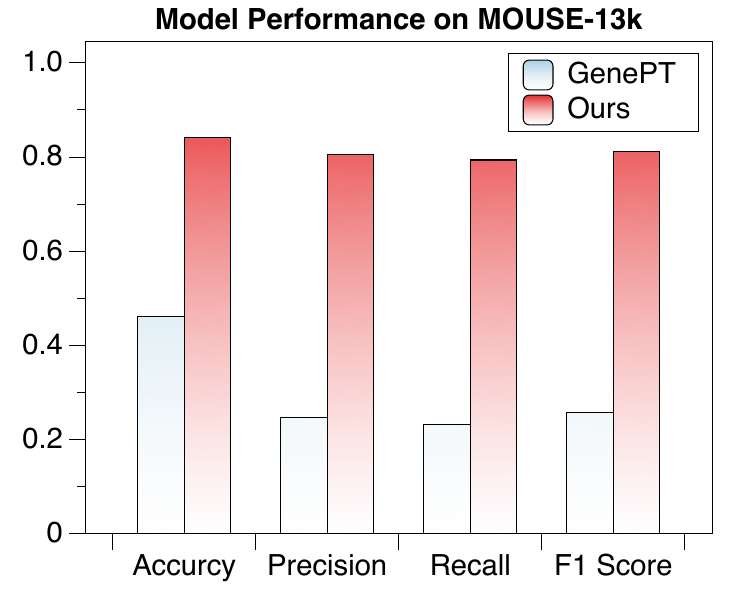}}
\caption{The performance comparison between \model\  and GenePT}
% \vspace{-0.5cm}
\label{fig:1}
\end{figure}

\noindent\textbf{\ul{Study of the Cell-type Annotation}.}
Figure~\ref{fig:1} reported the classification performance of GenePT and \model\ on the HUMAN-10k and MOUSE-13k. 
We used four classification metrics, accuracy, precision, recall, and F1 score to evaluate two methods.
We can observe that \model\ outperformed GenePT on two datasets with a huge margin.
The two methods compared have the same initial gene embedding, so we speculate that the common knowledge from the large language model could provide a better-supervised signal for the downstream task training, thus resulting in better performance. 
The equal performance among each metric from \model\ showed our proposed method and training strategy could provide robust downstream performance.

Additionally, we observed that, compared to the GenePT, our methods showed a more significant improvement on the HUMAN-10k dataset compared to the MOUSE-13k dataset. This may be due to the higher proportion of genes with NCBI knowledge entries in human genes (up to 97.5\%) compared to mouse genes (only 83.6\%). As a result, the embeddings of human cells benefited from better interpretation after processing with the large language model.

\begin{figure*}
    \centering
    \includegraphics[width=0.96\textwidth]{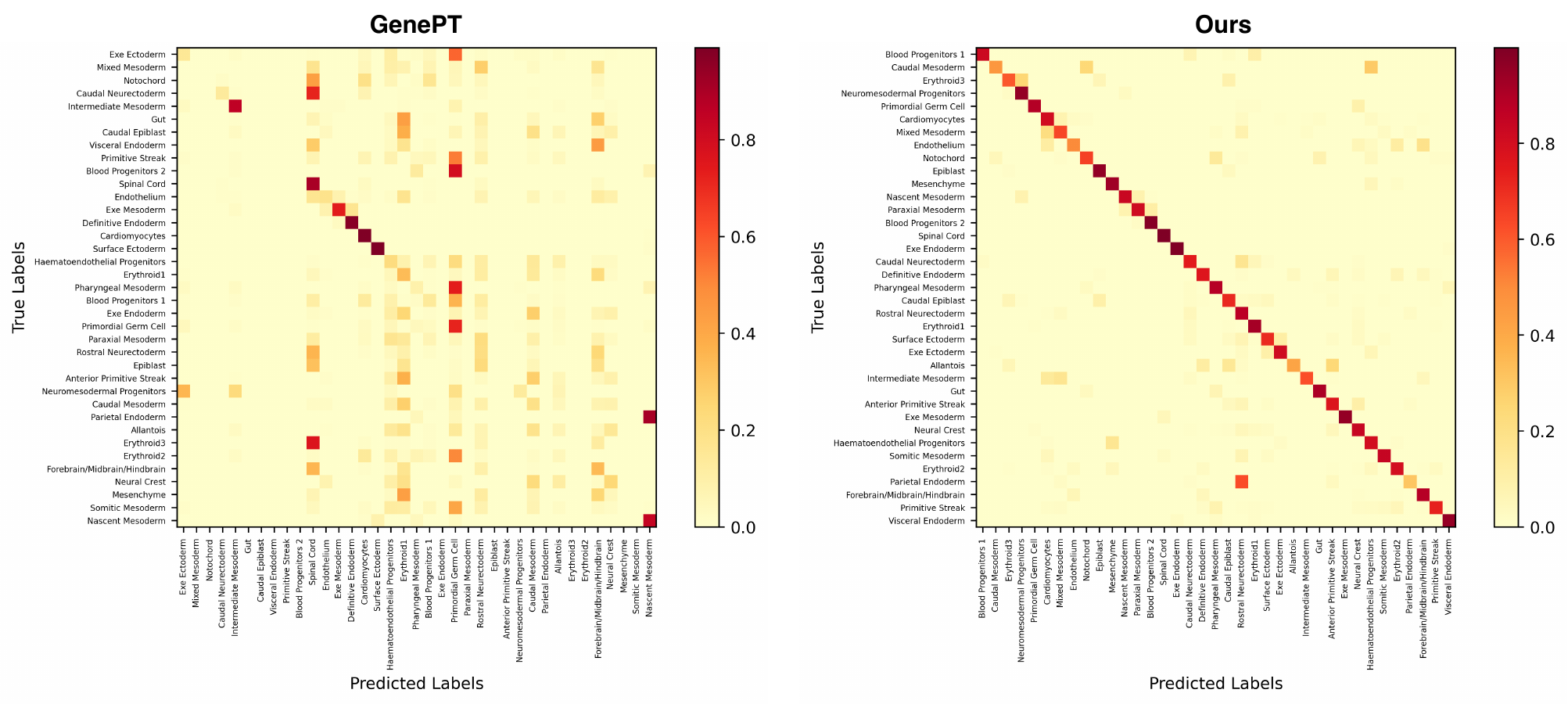}
    \caption{The confusion matrix of each method on MOUSE-13k.}
    % \vspace{-0.3cm}
    \label{fig:2}
\end{figure*}

\begin{figure*}[ht]
    \centering
    \begin{minipage}{0.46\textwidth}
        \centering
        \includegraphics[width=1\linewidth]{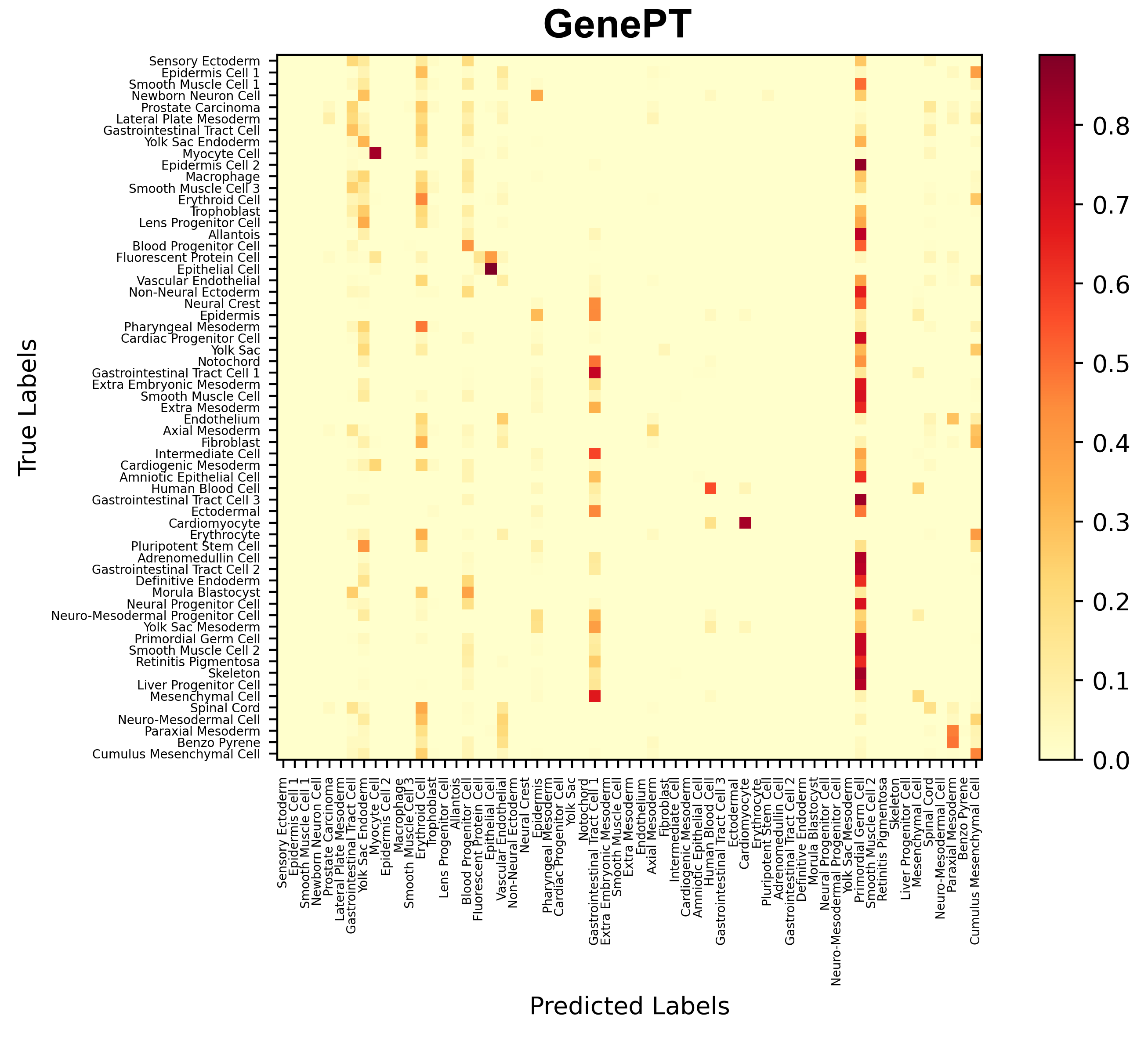}
    \end{minipage}
    \begin{minipage}{0.46\textwidth}
        \centering
        \includegraphics[width=1\linewidth]{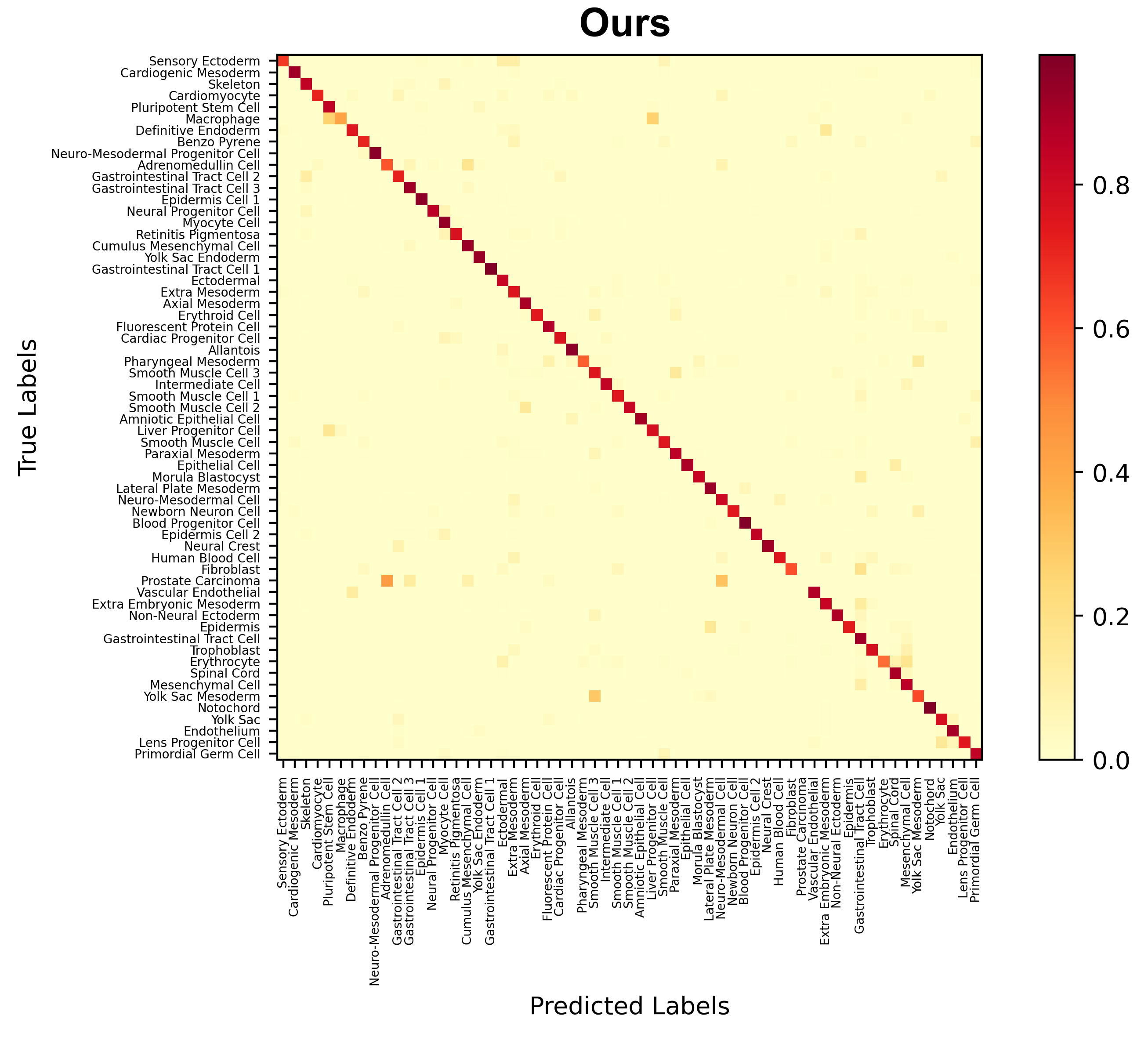}
    \end{minipage}
    \caption{The confusion matrix of each method on HUMAN-10k.}
    % \vspace{-0.3cm}
    \label{fig:3}
\end{figure*}

% \begin{figure}
%     \centering
%     \includegraphics[width=0.48\textwidth]{fig/mouse_com_matrix.pdf}
%     \caption{The confusion matrix of each method on HUMAN-10k.}
%     % \vspace{-0.3cm}
%     \label{fig:3}
% \end{figure}

% \begin{figure*}
%     \centering
%     \includegraphics[width=\textwidth]{fig/vis_mouse.pdf}
%     \caption{The UMAP illustration of the cell representation from initialization, GenePT, and \model\ on MOUSE-13k.}
%     % \vspace{0.2cm}
%     \label{fig:4}
% \end{figure*}

\begin{figure*}[ht]
    \centering
    \begin{minipage}{0.26\textwidth}
        \centering
        \includegraphics[width=1\linewidth]{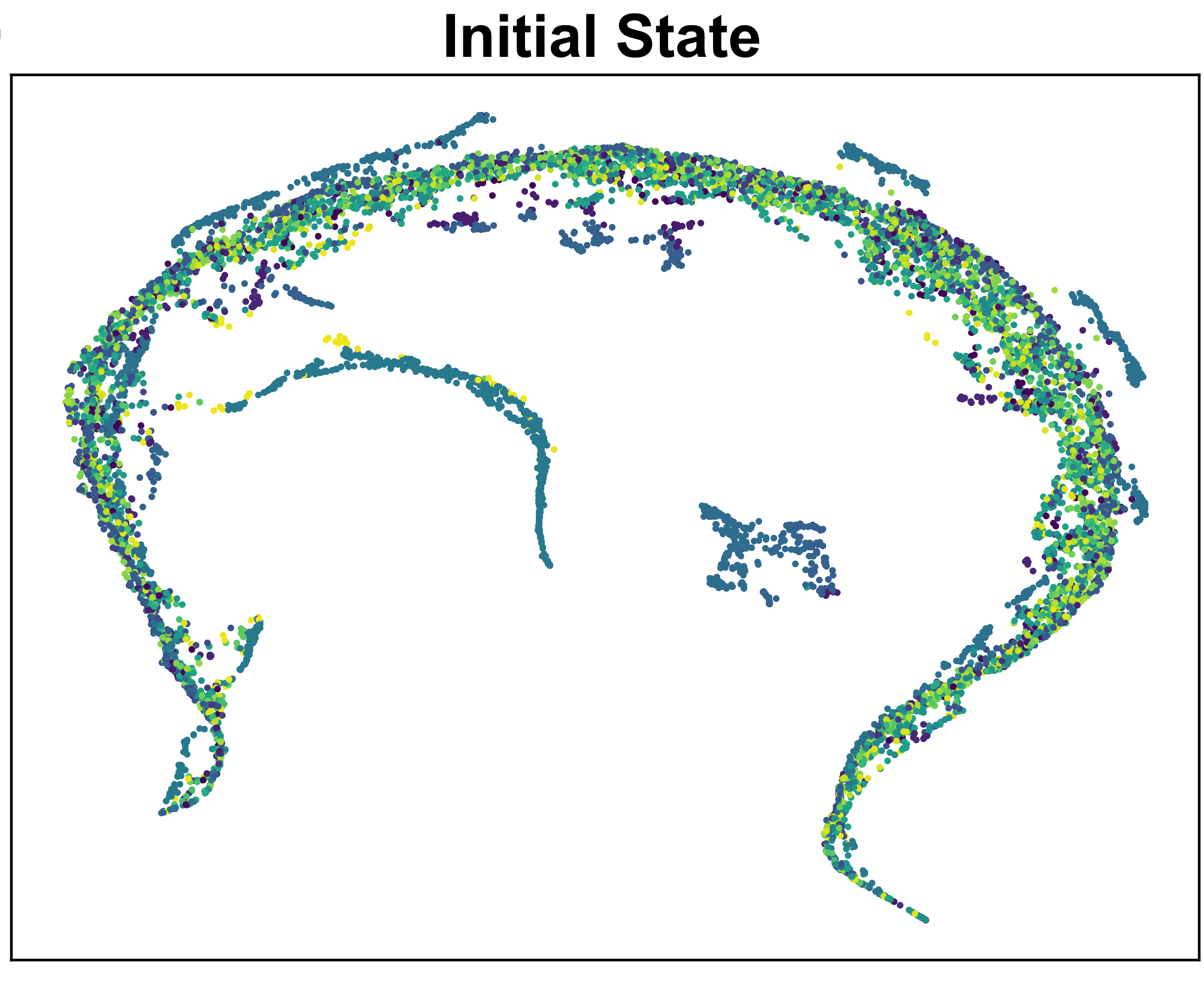}
    \end{minipage}
    \begin{minipage}{0.26\textwidth}
        \centering
        \includegraphics[width=1\linewidth]{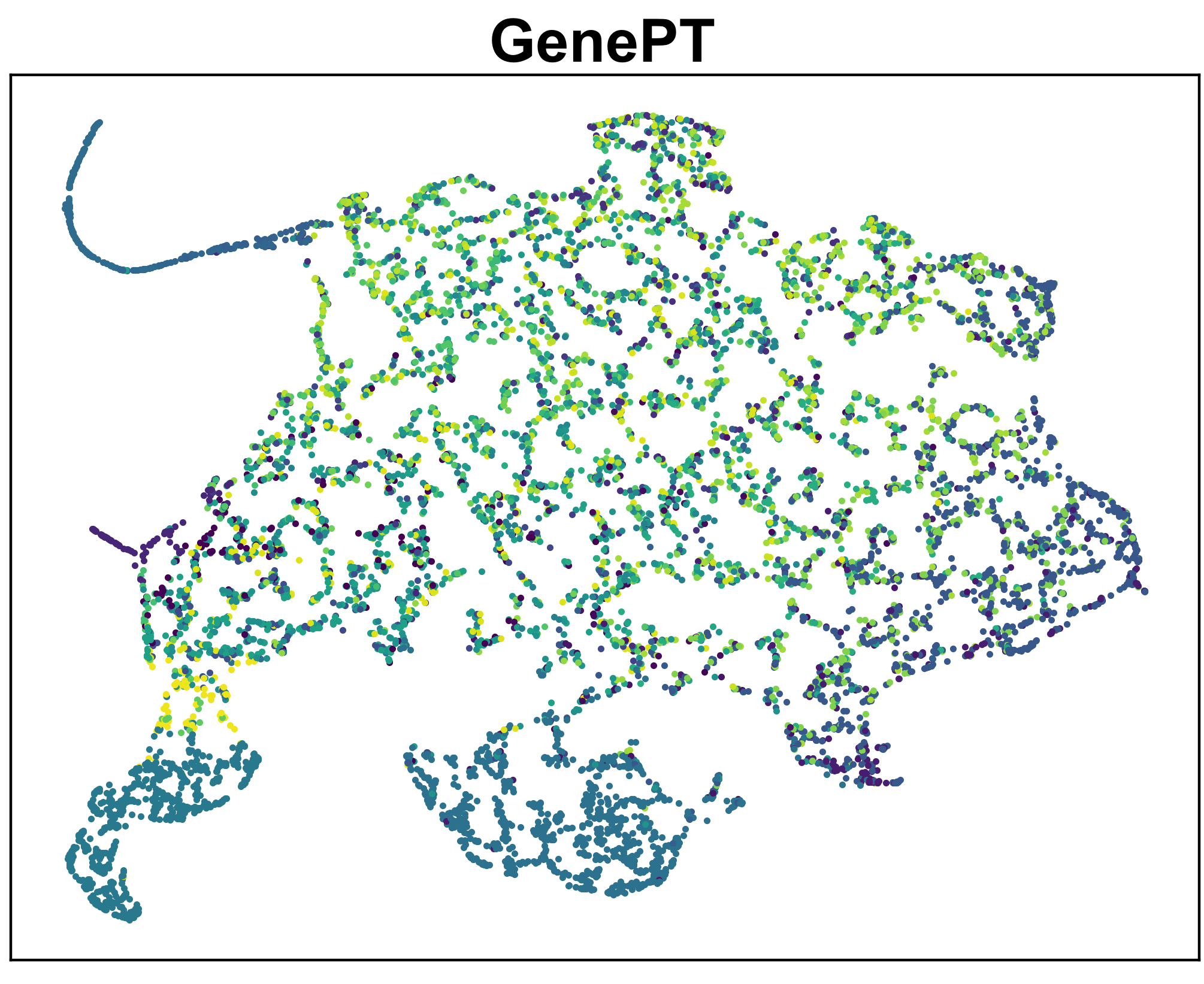}
    \end{minipage}
    \begin{minipage}{0.26\textwidth}
        \centering
        \includegraphics[width=1\linewidth]{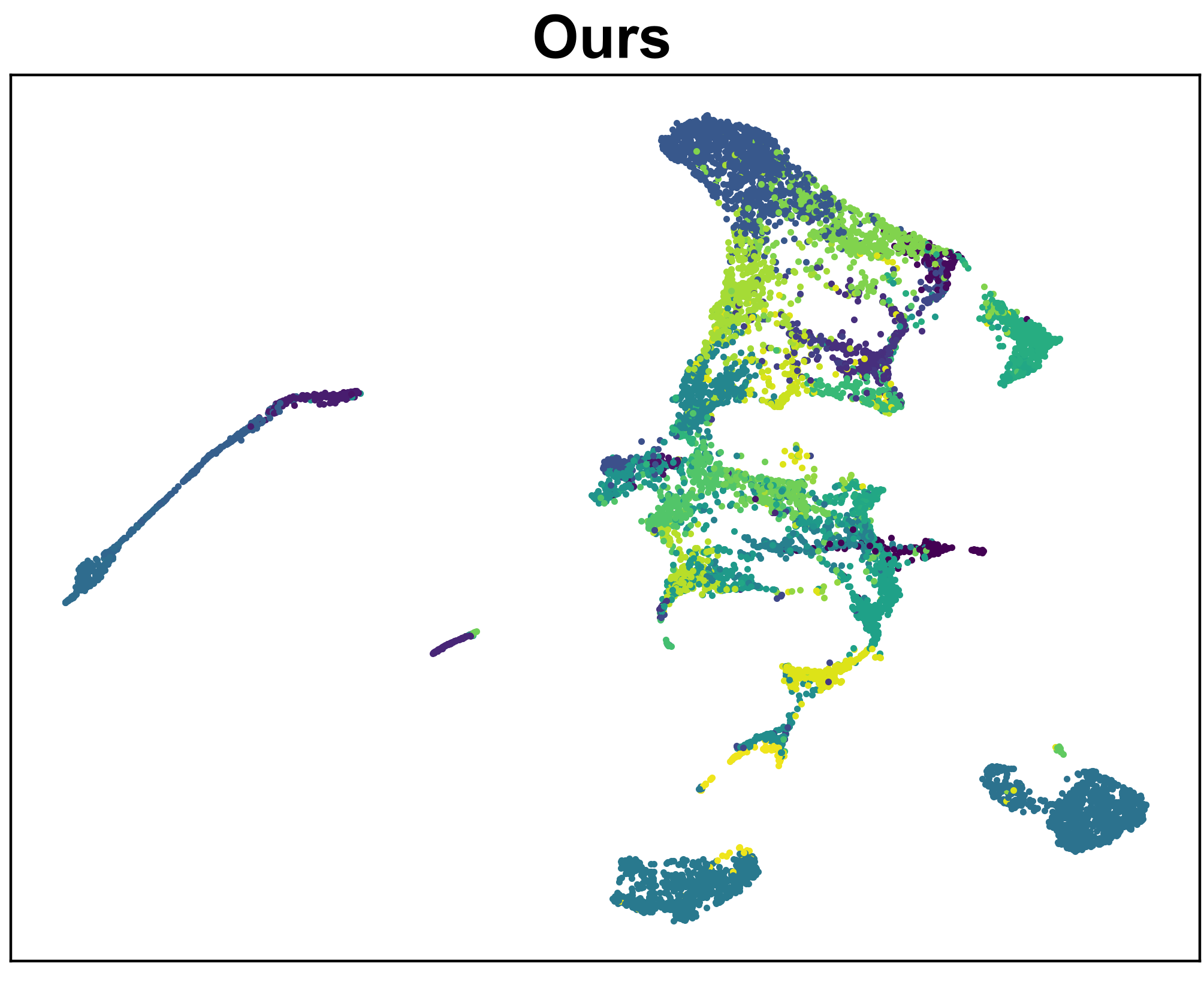}
    \end{minipage}
    \begin{minipage}{0.18\textwidth}
        \centering
        \includegraphics[width=1\linewidth]{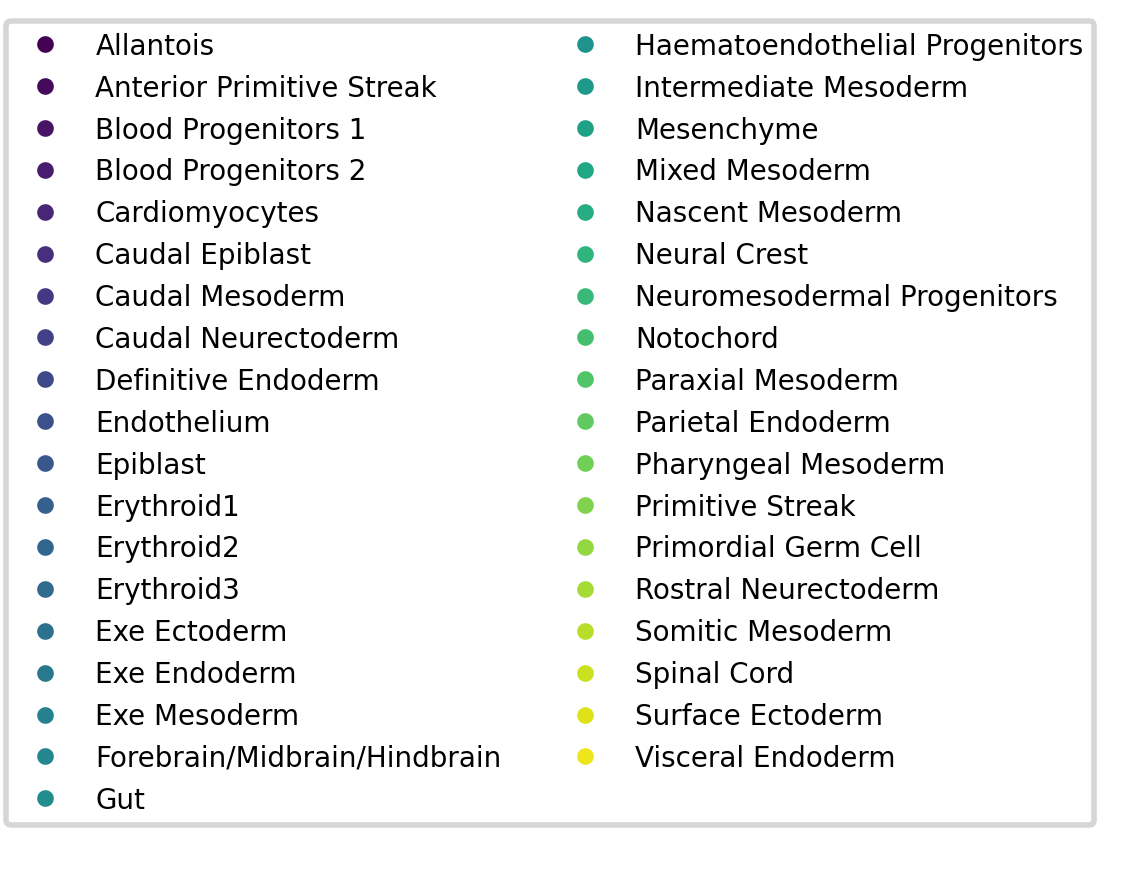}
    \end{minipage}
    \caption{The UMAP illustration of the cell representation from initialization, GenePT, and \model\ on MOUSE-13k.}
    % \vspace{0.2cm}
    \label{fig:4}
\end{figure*}

\begin{figure*}[ht]
    \centering
    \begin{minipage}{0.26\textwidth}
        \centering
        \includegraphics[width=1\linewidth]{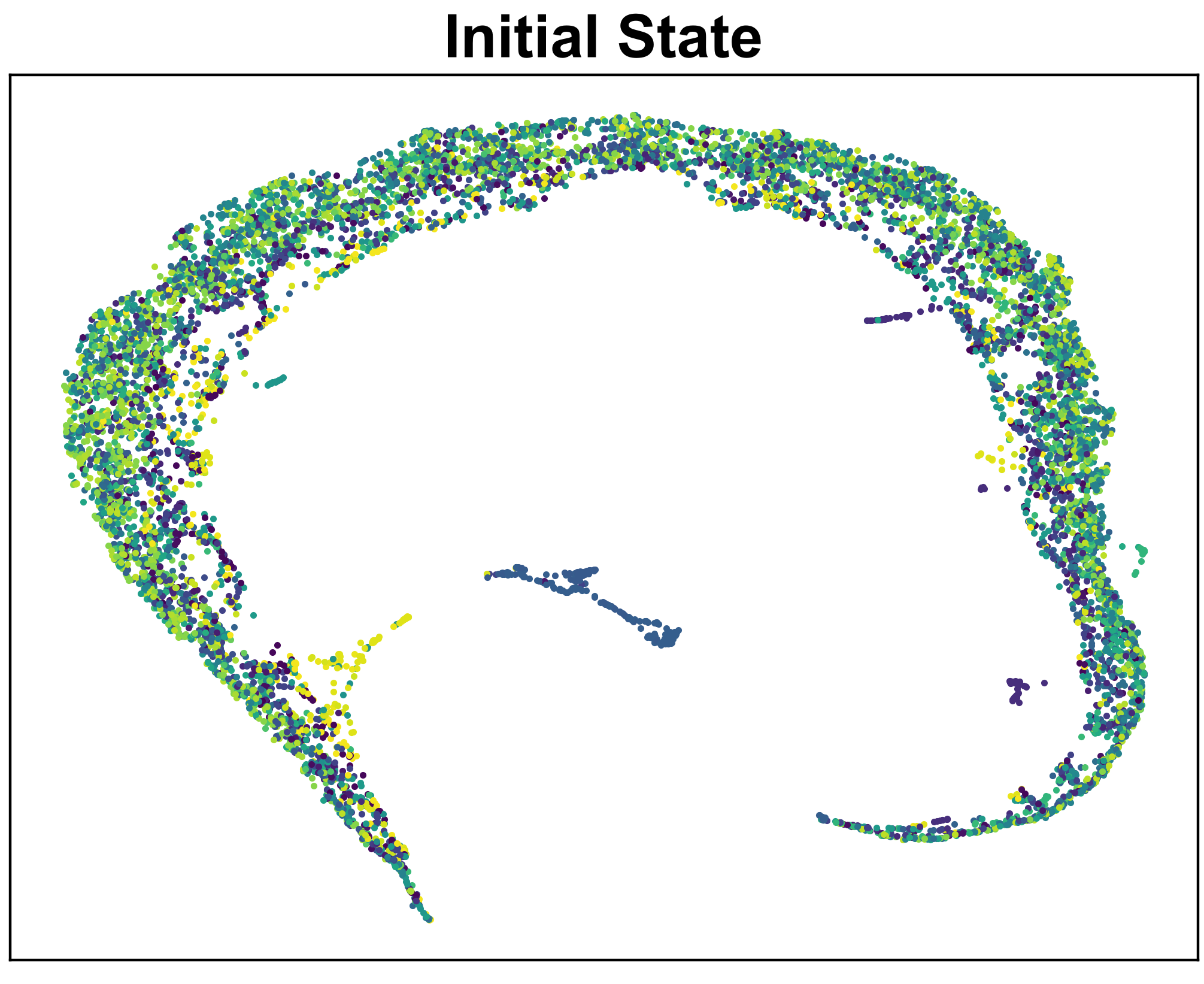}
    \end{minipage}
    \begin{minipage}{0.26\textwidth}
        \centering
        \includegraphics[width=1\linewidth]{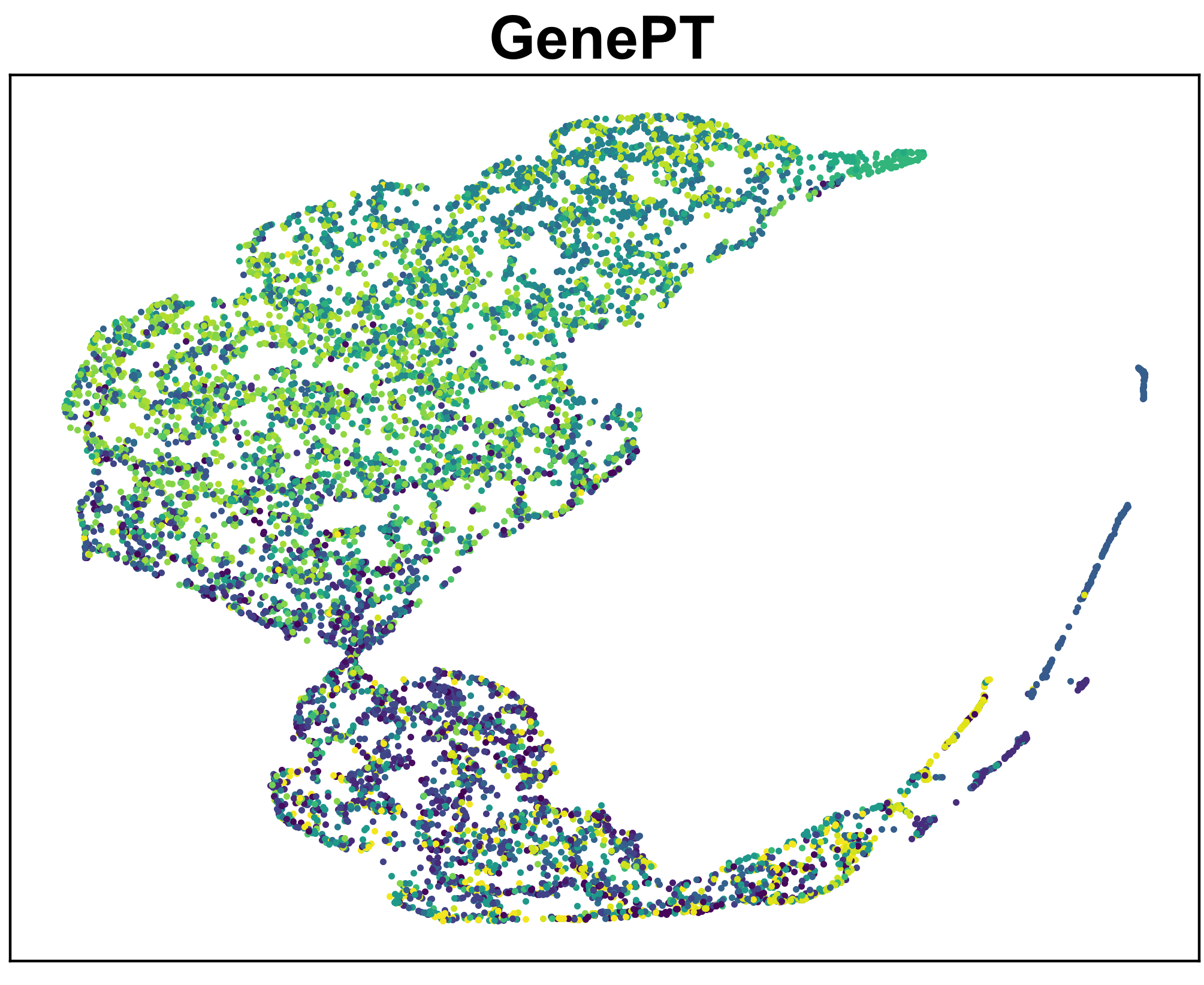}
    \end{minipage}
    \begin{minipage}{0.26\textwidth}
        \centering
        \includegraphics[width=1\linewidth]{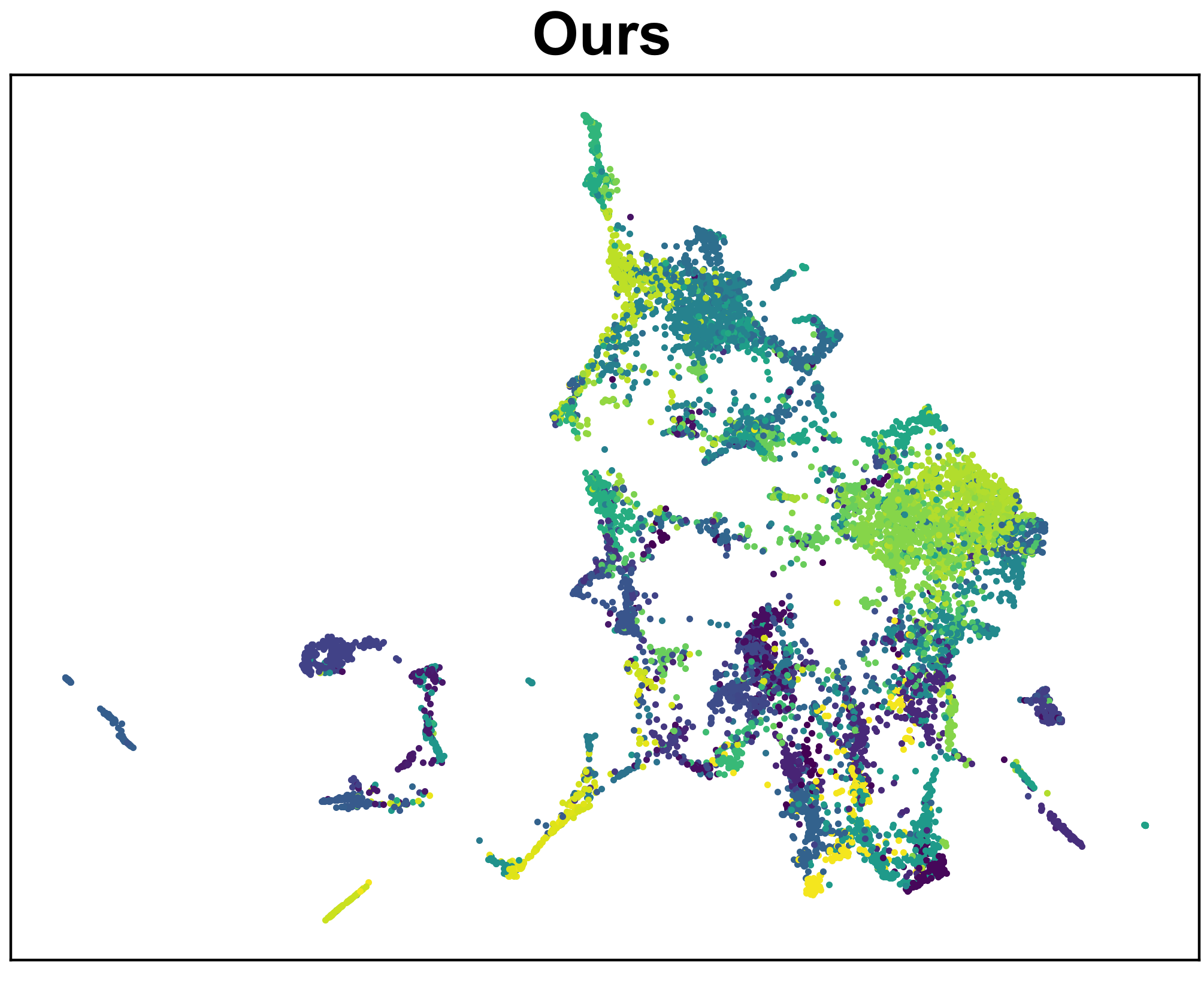}
    \end{minipage}
    \begin{minipage}{0.18\textwidth}
        \centering
        \includegraphics[width=1\linewidth]{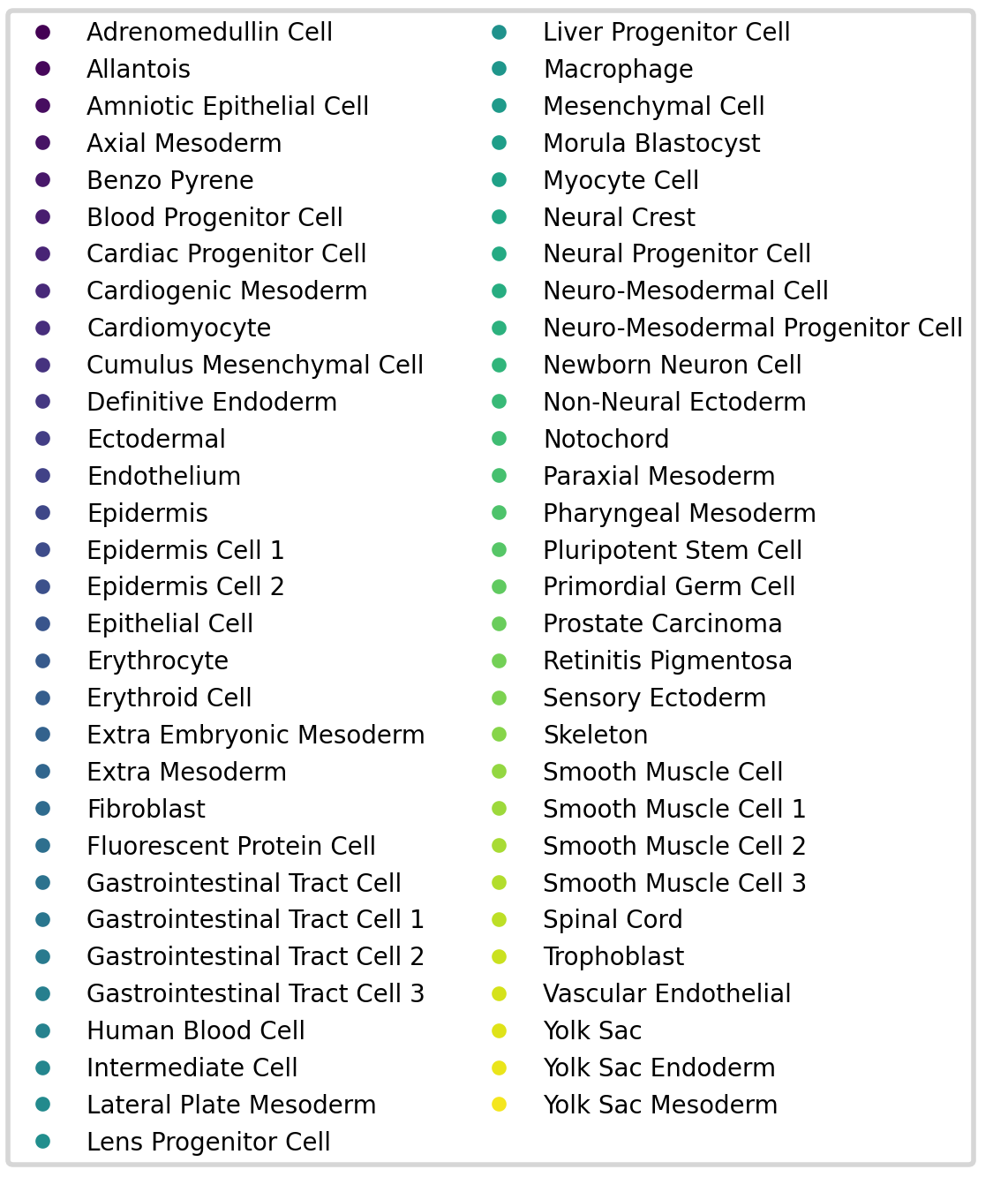}
    \end{minipage}
    \caption{The UMAP illustration of the cell representation from initialization, GenePT, and \model\ on HUMAN-10k.}
    \vspace{-0.3cm}
    \label{fig:5}
\end{figure*}

To step further, we reported the confusion matrix of each method on dataset MOUSE-13k, to show the difference between the prediction result and the true label. 
From the left part of Figure~\ref{fig:2}, we could observe that most prediction results were scattered across the matrix, indicating a high degree of misclassification. 
The lighter regions of the diagonal suggest that the model frequently confuses certain cell types with others, highlighting deficiencies in either feature representation or classification capability. 
In contrast, the right part of the matrix presents a starkly different scenario, where the majority of predictions align closely with the diagonal, thus reflecting a high concordance between predicted and actual labels. 
In addition to the overall performance improvement shown by the confusion matrix, we also observed that our method performs poorly on certain specific cell types, such as Parietal Endoderm. Our method tends to misclassify Parietal Endoderm cells as Rostral Neurectoderm, which are two significantly different cell types. Upon further investigation, we found that the MOUSE-13k dataset contains only 284 Parietal Endoderm cells, which is a very small number, while there are 5,392 Rostral Neurectoderm cells, making it the most numerous cell type in the dataset. This imbalance in data labels is likely the cause of the misclassification.

In Figure~\ref{fig:3}, we report the confusion matrix on the HUMAN-10k dataset. As can be seen, the confusion matrix on the left shows lighter and more chaotic colors along the diagonal, indicating significant misclassification of cell types. In contrast, the confusion matrix on the right, which represents our method, shows a notable improvement. The colors along the diagonal are darker, indicating that almost all cell types can be correctly identified.
This phenomenon denotes a significant improvement in classification accuracy, attributable to the sophisticated feature encoding and contextual understanding afforded by the prompt-based training method employed with the large language model.

\smallskip
\noindent\textbf{\ul{Study of the Visualization Analysis.}}
Figure~\ref{fig:4} reported the UMAP visualization of the cell embedding from the initial state, GenePT, and \model, on MOUSE-13k dataset. In the initial state of the MOUSE-13k datasets, as observed on the left side of the visualization, each cell types cluster together, exhibiting poor separability. 
The middle visualization, i.e. GenePT, shows some improvement in terms of separability among different categories. However, several cell types remain interspersed.
It can be observed that in the left clustering diagram, cell groups such as dark blue, yellow, and purple already exhibit noticeable cluster separation using only text-initialized embeddings without any training. In the middle UMAP plot, these cell groups still maintain good separation. While other cell groups show some improvement in separation, they do not exhibit a clear pattern. The clustering performance of the GenePT group is consistent with the results presented in the confusion matrix on the left side of Figure~\ref{fig:2}. The cell groups Definitive Endoderm, Cardiomyocytes, and surface ectoderm perform excellently in the confusion matrix, and their separation trends in the clustering diagram are more pronounced compared to other groups.
In contrast, the visualization of \model{} demonstrates a markedly superior clustering effect. 
Cells of the same type exhibit exceptional aggregative properties, suggesting a high degree of intra-class similarity and inter-class divergence. 
Additionally, the visualization of \model{} demonstrates improvements corresponding to the GenePT group. 
Cell types such as Definitive Endoderm, Cardiomyocytes, and surface ectoderm maintain their original excellent performance.
Other previously mixed groups, such as the Nascent Mesoderm, Primitive Streak, and Spinal Cord, which were mixed together in the GenePT group, show better separation trends in the right visualization. 
This improvement is also reflected in the right-side confusion matrix of Figure~\ref{fig:2}.
% This distinction becomes more pronounced when considering that both GenePT and our method utilize a comparable number of trainable parameters and employ identical labeling schemes for training.

In Figure~\ref{fig:5}, on the HUMAN-10k dataset, our method also shows significant improvement in transcriptome interpretation. The left clustering diagram shows that the human-initialized cell embeddings tend to be mixed together. The middle GenePT clustering diagram shows a slight improvement; previously well-separated cells, such as Epithelial cells and Cardiomyocytes, remain clearly separated, while other groups show only slight improvement. This aligns with the left-side confusion matrix in Figure~\ref{fig:3}, where these cell types perform better than others. In the right UMAP plot, cells of the same type, represented by the same color, show more organized clustering, demonstrating a substantial improvement. This indicates that our method using LLM for transcriptome interpretation is quite effective.
The underlying driver is that our proposed method leverages the broad, generalized knowledge inherent in LLMs to provide more effective supervisory signals, thereby enhancing the separability of learned cell embeddings.
% Such a result not only underscores the efficacy of incorporating LLMs into cell embedding but also illuminates the potential for direct interpretation of genetic sequences by these models.

\section{Conclusion Remarks}
This study represents a preliminary stride in the application of leveraging large language models for interpreting single-cell omics data, particularly in the context of cell-type annotation. 
The experimental results demonstrate that integrating LLMs into single-cell omics analysis pipelines can significantly enhance our ability to interpret and classify cell types. 
Additionally, extending this method to multi-omics integration and rare cell type identification could yield valuable insights for precision medicine and developmental biology.
In conclusion, our study marks a significant step forward in the application of artificial intelligence to single-cell biology, paving the way for more sophisticated, knowledge-driven approaches to understanding cellular complexity at unprecedented resolution.
% The integration of LLMs into the analysis of gene expression data opens new frontiers in computational biology, offering novel methods to decode complex biological information and paving the way for groundbreaking discoveries in the field. 

\clearpage
\balance
\bibliographystyle{IEEEtran}
%\vspace{-0.3cm}
\bibliography{ref}

\end{document}